\documentclass[%
 reprint,
 amsmath,amssymb,
 aps
]{revtex4-1}

\usepackage{graphicx}
\usepackage{dcolumn}
\usepackage{bm}

\usepackage{eufrak}

\usepackage{aurical}
\usepackage[T1]{fontenc}
\usepackage{vicent}
\usepackage[OT1]{fontenc}

\usepackage[version=3]{mhchem}
\usepackage{breakurl}

\begin{document}


\title{Excitation and ionisation cross-sections in condensed-phase \mbox{biomaterials} by electrons down to very low energy:
application to liquid water and genetic building blocks}

\author{Pablo de Vera$^1$}
 \email{pablo.vera@um.es}

\author{Isabel Abril$^2$} 

\author{Rafael Garcia-Molina$^1$}


\affiliation{%
 $^1$Departamento de Física -- Centro de Investgación en Óptica y Nanofísica, Universidad de Murcia, Murcia
}%
\affiliation{%
 $^2$Departament de Física Aplicada, Universitat d'Alacant, Alacant, Spain
}%


\date{\today}

\begin{abstract}
Electronic excitations and ionisations produced by electron impact are key processes in the radiation-induced damage mechanisms in materials of biological relevance, underlying important medical and technological applications, including radiotherapy, radiation protection in manned space missions and nanodevice fabrication techniques. However, experimentally measuring all the necessary electronic interaction cross-sections for every relevant material is an arduous task, so it is necessary having predictive models, sufficiently accurate yet easily implementable. 
In this work we present a model, based on the dielectric formalism, to provide reliable ionisation and excitation cross-sections for electron-impact on complex biomolecular media, taking into account their condensed-phase nature. We account for the indistinguishability and exchange between the primary beam and excited electrons, for the molecular electronic structure effects in the electron binding, as well as for higher-order corrections to the first Born approximation. The resulting approach yields total ionisation cross-sections, energy distributions of secondary electrons, and total electronic excitation cross-sections for condensed-phase biomaterials, once the electronic excitation spectrum is known, either from experiments or from a predictive model. The results of this methodology are compared with the available experimental data in water and DNA/RNA molecular building blocks, showing a very good agreement and a great predictive power in a wide range of electron incident energies, from the large values characteristic of electron beams down to excitation threshold. The proposed model constitutes a very useful procedure for computing the electronic interaction cross-sections for arbitrary biological materials in a wide range of electron incident energies.
\end{abstract}

\maketitle


\section{Introduction}

Charged particles interacting with matter lose their energy mainly through electronic excitations and ionisations, which result in the ejection of secondary electrons and the ensuing generation of electron cascades \cite{Nikjoo2012}, mostly having energies below 100 eV. Such electrons (even those with energies below ionisation threshold) are capable of inducing very harmful effects in organic and biological matter, either by means of electronic excitations leading to molecular fragmentation or by dissociative electron attachment \cite{Boudaiffa2000,Thorman2015}. The characteristics of the electron cascade when reaching target sensitive molecules (e.g. DNA building blocks or polymeric lithographic resists) can be obtained through detailed Monte Carlo simulations \cite{Nikjoo2012,Dapor2017,Incerti2018,Schuemann2019,DaporBook}, the reliability of whose predictions strongly depends on the accuracy of the cross-sections (i.e., interaction probabilities) with which they are fed for simulating the generation and propagation of these electrons through condensed matter. This knowledge is essential for the better understanding, through modelling, of numerous medical and technological applications, including radiation therapy for cancer \cite{Schardt2010,Solovyov2017} or advanced nanofabrication techniques \cite{Huth2018,deVera2020}.

Plenty of experimental information is available on electron-impact ionisation cross-sections (total probabilities as well as energy spectra of ejected electrons) of relevant biomolecules (such as water or DNA nucleobases), however it is usually limited to the gas phase \cite{Bolorizadeh1986,Itikawa2005,Fuss2009,Bull2014,Rahman2016,Bug2017,Wolff2019}, without taking into account the condensed phase nature of living organisms. Experimental data on electronic excitations are much scarcer and scattered \cite{Thorn2007,Brunger2008,Thorn2008,Matsui2016,Bug2017} and also limited just to molecules in the gas phase, with only a few exceptions \cite{Michaud2012,Lemelin2016}. An alternative for gathering all relevant cross-sections resorts into theoretical calculations, where some very successful \textit{ab initio} approaches are available for calculating both excitation and ionisation cross-sections \cite{Bouchiha2006,Gorfinkiel2020}. However, the required tools and background are usually rather complex and not accessible for a wide audience. There are also a number of very popular semiempirical models \cite{ICRU55,Mozejko2003,Mozejko2005}, which are easy to implement but, in return, remain generally limited to the ionisation process and/or they cannot provide the energy spectrum of the generated electrons, which is essential for following up their transport in the simulations.

A convenient alternative for studying ion-impact ionisation of condensed-phase organic and biological materials, based on the dielectric formalism \cite{deVera2013,deVera2015}, has the advantage that (i) it is relatively simple to implement, (ii) it provides both the total cross-sections and the energy and angular spectra of secondary electrons, and (iii) it only requires a few easily accessible parameters of the organic target (namely, their optical properties or just their atomic composition and density, together with their outer-shell ionisation thresholds). Thus, this procedure can be straightforwardly applied to any complex biological material, including liquid water, RNA/DNA and their building blocks, proteins \cite{deVera2013b} or even subcellular compartments \cite{deVera2014}. The extension of this approach for electron projectiles requires accounting for a series of improvements. First, the indistinguishability and exchange between the primary and the struck electrons need to be appropriately considered both for electronic excitations and ionisations \cite{deVera2019}. Second, a series of low energy modifications are necessary: (i) higher-order corrections \cite{Emfietzoglou2005,deVera2019} to the perturbative nature of the first Born approximation (on
which the dielectric formalism holds), and (ii) avoid counting ionisations attributed to low energy electrons that cannot ionise all the electronic levels of the target molecules \cite{Tan2018}.

In this work, we extend the dielectric response model \cite{deVera2013,deVera2015} to electrons in a wide energy range, covering from the high energies typical of electron beams or delta-electrons down to the low energies characteristic of hot electrons. This extension introduces corrections to the first Born approximation to describe very low energy electrons, and includes the indistinguishability and exchange, considering how electronic excitations and ionisations are differently affected by the target electronic structure.

The resulting methodology yields the total ionisation and excitation cross-sections, as well as the energy distribution of secondary electrons. Calculations for liquid water, all the DNA/RNA bases (adenine, thymine, guanine, cytosine and uracil) and tetrahydrofuran (THF, an important analogue molecule of the sugar component of the phosphate-deoxyribose backbone of DNA), in the condensed phase, are presented and compared to available experimental data, showing an excellent general agreement.

The model presented in this work provides a convenient and universal approach for reliably evaluating the main quantities involved in electron transport and effects within arbitrary organic and biological materials (accounting for their condensed-phase state) in a wide energy range. The methodology and data presented in this work will result very useful in the field of ionising radiation interaction with matter, either for medical or technological applications.

The work is organized as follows. In Section \ref{sec:theory} we provide the theoretical background and introduce the necessary improvements, whose results are analysed and compared with available experimental data in Section \ref{sec:results}. Finally, we present in Section \ref{sec:concl} a summary and the conclusions of our work.

\section{Theoretical framework}
\label{sec:theory}

\subsection{Electronic excitation and ionisation of biomaterials by swift charged particle impact}
\label{sec:drf}

The doubly differential cross-section (DDCS) $\frac{{\rm d}^2 \sigma(T,E,k)}{{\rm d}E {\rm d}k}$ for the scattering of a charged particle, having kinetic energy $T$, with an atom or molecule gives the probability for the incident particle to lose some specific amount of energy $E$ and of momentum $\hbar k$ as a result of the interaction. In the quantum theory of scattering, the DDCS can be related to the direct $f(\Vec{k_1},\Vec{k_2})$ and the exchange $g(\Vec{k_1},\Vec{k_2})$ scattering amplitudes \cite{Prasad1965, Rudge1968}:
\begin{equation}
\frac{{\rm d}^2 \sigma(T,E,k)}{{\rm d}E {\rm d}k} \propto |f(\Vec{k_1},\Vec{k_2})|^2+|g(\Vec{k_1},\Vec{k_2})|^2-{\rm Re}[f(\Vec{k_1},\Vec{k_2}) g^*(\Vec{k_1},\Vec{k_2})] \, \mbox{,}
    \label{eq:DDCS1}
\end{equation}
where $\hbar \Vec{k_1}$ and $\hbar \Vec{k_2}$ are, respectively, the momentum of the scattered and excited particles in the final state. The last term in Eq. (\ref{eq:DDCS1}) accounts for the interference between the direct and exchange scattering amplitudes.

An incident ion and an excited electron are distinguishable particles, so the exchange and interference terms vanish in Eq. (\ref{eq:DDCS1}). In such a case, the first Born approximation (FBA) relates the DDCS only to the direct scattering amplitude, $\left.\frac{{\rm d}^2 \sigma(T,E,k)}{{\rm d}E {\rm d}k}\right|_{\rm FBA} \propto |f(\Vec{k_1},\Vec{k_2})|^2$, and the dielectric formalism provides a compact expression for the DDCS for a charged particle scattered in a condensed-phase medium \cite{Lindhard1954, Nikjoo2012}. This is valid when the energy loss is small in comparison with the energy of the primary particle, i.e., for incident ions or electrons much faster than the target electrons, and it is given by:
\begin{equation}
    \left.\frac{{\rm d}^2 \sigma(T,E,k)}{{\rm d}E {\rm d}k}\right|_{\rm FBA} = \frac{e^2 [Z-\rho(k)]^2}{\pi \hbar^2 {\cal N}} \frac{M}{T} \frac{1}{k} {\rm Im}\left[ \frac{-1}{\epsilon(k,E)} \right] \, \mbox{,}
    \label{eq:FBAions}
\end{equation}
where $M$ is the incident particle mass, $Z$ is its charge, and $\rho(k)$ is the Fourier transform of its charge density. For electron projectiles, $Z=1$, $\rho(k)=1$ (since they are point charges) and $M=m$ is the electron mass. The so-called energy-loss function (ELF), ${\rm Im}\left[ \frac{-1}{\epsilon(k,E)} \right]$, represents the electronic excitation spectrum of the target material, with $\epsilon(k,E)=\epsilon_1(k,E)+i\epsilon_2(k,E)$ being the complex dielectric function. ${\cal N}$ is the target atomic (or molecular) density, which relates the microscopic $\sigma$ and the macroscopic cross-section $\Lambda$ (or inverse mean free path, IMFP) as $\Lambda = \lambda^{-1} = {\cal N} \sigma$, with $\lambda$ being the electronic inelastic mean free path.

The ELF of the target material accounts for the excitation and ionisation spectrum of its outer- and inner-shell electrons:
\begin{equation}
    {\rm Im}\left[ \frac{-1}{\epsilon(k,E)} \right] = {\rm Im}\left[ \frac{-1}{\epsilon(k,E)} \right]_{\rm out} + \sum_j {\rm Im}\left[ \frac{-1}{\epsilon(k,E)} \right]_j \, \mbox{,}
    \label{eq:ELF}
\end{equation}
where the sum goes over all the $j$-inner shells of the atoms forming the material, whose ELF can be straightforwardly calculated by means of hydrogenic generalized oscillator strengths \cite{Heredia-Avalos2005}. As for the outer-shell electrons, the ELF is usually obtained from optical ($\hbar k=0$) experiments. Its structure is rather similar for most biological and organic materials, presenting a main broad excitation peak around 20 eV, and thus it can be conveniently described by means of a single-Drude function, whose parameters can be determined as a function of the material composition and density \cite{Tan2004, deVera2013, deVera2013b}. This approach allows obtaining a reliable excitation and ionisation spectrum for any arbitrary condensed organic material, irrespectively of whether it has been measured or not. The optical ELF is extended to finite momentum transfers ($\hbar k\neq 0$) using the Mermin Energy-Loss Function – Generalized Oscillator Strengths (MELF-GOS) methodology \cite{Heredia-Avalos2005}, which has proven to be very successful for handling condensed-phase materials \cite{Heredia-Avalos2005, Heredia-Avalos2007d,Denton2008,Denton2008a,deVera2014a, deVera2019}, and particularly liquid water \cite{Garcia-Molina2009, Garcia-Molina2011}. The main physical properties of the biomaterials studied in this work (needed for further calculations) are summarized in Table \ref{tab:properties}.

\begin{table*}
\small
  \caption{\ Physical properties of the biomaterials discussed in this work. 
  }
  \label{tab:properties}
  \begin{tabular*}{\textwidth}{@{\extracolsep{\fill}}llllllll}
    \hline
    Target & Liquid water & THF & Uracil & Adenine & Guanine  & Cytosine & Thymine\\
    \hline
    Chemical formula & \ce{H2O} & \ce{C4H8O} & \ce{C4H4O2N2} & \ce{C5H5N5} & \ce{C5H5ON5}  & \ce{C4H5ON3} & \ce{C5H6O2N2}\\
    Atomic number $Z$ & 10 & 40 & 58 & 70 & 78 & 58 & 66 \\
    Atomic mass $A$ & 18.0 & 72.11 & 112.09 & 135.13 & 151.13 & 110.10 & 126.11\\
    Density (g/cm$^3$) & 1.00 & 0.89 \cite{ChemSynthesis} & 1.40 \cite{Isaacson1972} & 1.35 \cite{Tan2004} & 1.58 \cite{Tan2004} & 1.30 \cite{Tan2004} & 1.36 \cite{Isaacson1972}\\
    First binding energy $B_1$ (eV) & 10.79 \cite{Dingfelder1998} & 9.74 \cite{Bug2017} & 9.50 \cite{Galassi2012} & 8.44 \cite{Bernhardt2003} & 8.24 \cite{Bernhardt2003} & 8.94 \cite{Bernhardt2003} & 9.14 \cite{Bernhardt2003}\\
    Last binding energy $B_n$ (eV) & 32.30 \cite{Dingfelder1998} & 36.97 \cite{Bug2017} & 37.92 \cite{Galassi2012} & 37.48 \cite{Bernhardt2003} & 39.25 \cite{Bernhardt2003} & 37.70 \cite{Bernhardt2003} & 37.81 \cite{Bernhardt2003}\\
    Mean binding energy $B_{\rm mean}$ (eV) & 13.71 
    & 15.12 
    & 15.40 
    & 14.66 
    & 15.17 
    & 14.76 
    & 15.28 
    \\
    Excitation threshold $E_{\rm th}$ (eV) & 7.00 \cite{Grand1979} & 7.94 \cite{Dwivedi2015} & 3.65 \cite{Nguyen2004} & 4.47 \cite{Silaghi2005} & 4.31 \cite{Silaghi2005} & 4.46 \cite{Silaghi2005} & 4.54  
     \cite{Yamada1968}\\
    Optical ELF & Exp. \cite{Hayashi2000} & Param. model \cite{Tan2004} & Exp. \cite{Isaacson1972} & Exp. \cite{Arakawa1986} & Exp. \cite{Emerson1975} & Param. model \cite{Tan2004} & Exp. \cite{Isaacson1972}\\
    Num. valence electrons $N_{\rm val}$ & 8 & 30 & 42 & 50 & 56 & 42 & 48\\
    Num. effective electrons $N_{\rm eff}$ \cite{Mendez2020} & 6 & 28 & 36 & 45 & 49 & 37 & 42\\
    \hline
  \end{tabular*}
\end{table*}

\begin{figure*}
 \centering
 \includegraphics[width=0.8\textwidth]{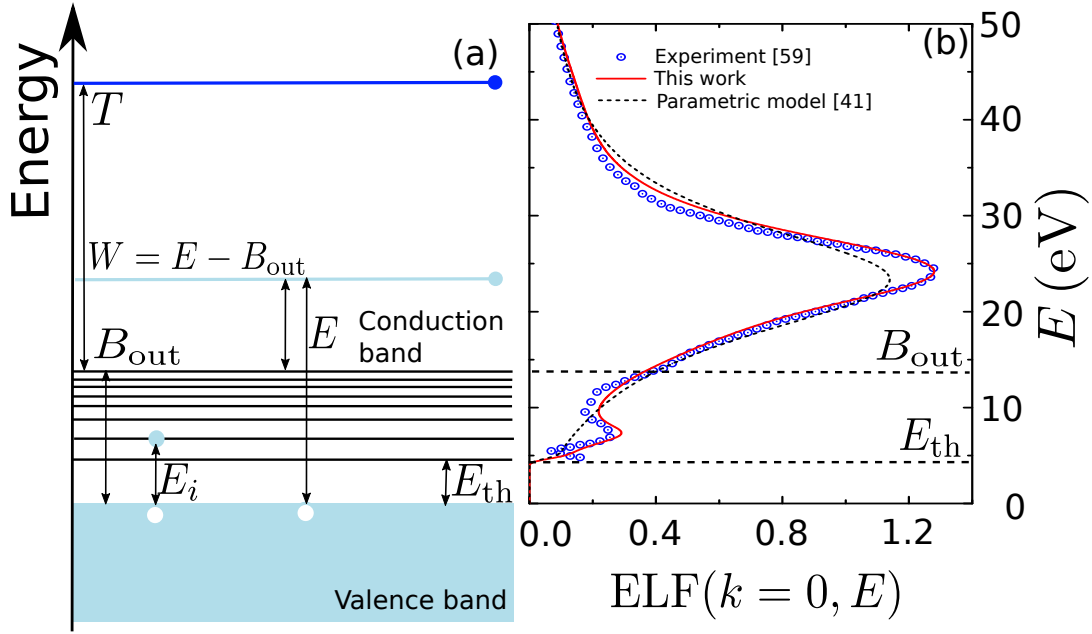}
 \caption{(a) Diagram of the possible electronic transitions in an insulator. Between the valence and the conduction bands there are localized states that can be reached when electrons in the valence band gain an energy $E_i$; see main text for details. (b) Optical energy loss-function of adenine as a function of the energy transfer $E$. Circles are experimental data for solid adenine \cite{Arakawa1986}, the solid line represents the MELF-GOS calculation \cite{Heredia-Avalos2005}, whereas the dashed line corresponds to the predictive parametric model \cite{Tan2004} (with a threshold energy $E_{\rm th}$).}
 \label{fig:ELF}
\end{figure*}

Figure \ref{fig:ELF}(a) illustrates schematically the electronic transitions in an insulating target (which is the case for biological matter). The outer-shell electrons of the material occupy the valence band, formed by localized states in the target molecules, while the conduction band, which hosts the mobile electrons, initially only contains the primary electron, moving with kinetic energy $T$. Between both bands, there are empty discrete states, to which electrons from the valence band can be promoted by excitation after receiving an energy transfer $E_i$. The processes of electronic excitation and ionisation are all contained in the ELF of the biological target, and they cannot be easily disentangled, except for some specific cases such as water \cite{Dingfelder1998,Emfietzoglou2003b, Emfietzoglou2005, Emfietzoglou2007c,Emfietzoglou2007a}. As an illustration of a typical biomaterial, Fig. \ref{fig:ELF}(b) depicts the experimental optical ELF of adenine from 0 to 50 eV (circles),\cite{Arakawa1986} together with the prediction of the above mentioned parametric predictive model for biomaterials (dashed line) \cite{Tan2004} and the fitting to the experimental data by means of the MELF-GOS method (solid line). \cite{Heredia-Avalos2005}   Electronic excitations start to appear for energy transfers $E$ greater or equal than the first excitation threshold $E_{\rm th}$, while ionisation processes will occur for larger energy transfers. Although the parametric model\cite{Tan2004} does not account for the excitation threshold, we applied a cutoff to the single-Drude function below the corresponding value (see Table \ref{tab:properties}).

When considering ion-impact ionisation \cite{deVera2013, deVera2015} we adopted some simplifying assumptions in order to disentangle the excitation and ionisation processes for arbitrary biomaterials. Since electronic excitations are mostly restricted to low energies, while most of the larger energy transfers produce ionisation, we defined a mean binding energy $B_{\rm out}$ for the outer shell electrons, obtained as a direct average of the target binding energies, which can be found either from experiment or from \textit{ab initio} calculations. In this approach, it was assumed that any energy transfer $E>B_{\rm out}$ produces a secondary electron with kinetic energy $W=E-B_{\rm out}$ measured from the bottom of the conduction band (see Fig. \ref{fig:ELF}(a)). Similarly, secondary electrons with energy $W=E-B_j$ can be produced by ionisation of the inner shells, where $B_j$ are the corresponding binding energies. In contrast, any energy transfer $E<B_{\rm out}$ is considered to be an electronic excitation, to some discrete and localized energy level through an energy transfer $E_i$, without electron ejection to the conduction band. This approximation for ion impact produced very good results in comparison with experimental data \cite{deVera2013}. 

The DDCS, Eq. (\ref{eq:FBAions}), can be used to obtain the angular distribution of secondary electrons by relating the momentum transfer to the ejection angle \cite{deVera2015}. It can also be integrated in momentum transfer in order to yield the energy distribution of secondary electrons, as well as in energy to obtain the total probabilities for ionisation and excitation \cite{deVera2013}. While the integration limits are straightforwardly deduced for the case of ion impact \cite{deVera2013, deVera2015}, a more detailed analysis of the possible electronic transitions is needed in the case of electron impact due to indistinguishability \cite{deVera2019}, for which the sketch of Fig. \ref{fig:ELF}(a) is useful. The singly differential ionisation cross-section (SDCS) for electron impact is obtained as:
\begin{widetext}
\begin{equation}
    \left.\frac{{\rm d}\sigma^{\rm ionis}(T,W)}{{\rm d}W}\right|_{\rm FBA} = 
    \frac{e^2}{\pi \hbar^2 {\cal N}} \frac{m}{T} \left\{ 
    \int_{k_{-,{\rm out}}}^{k_{+,{\rm out}}} \frac{{\rm d}k}{k} {\rm Im}\left[ \frac{-1}{\epsilon(k,W+B_{\rm out})} \right]_{\rm out} 
    + 
    \sum_j \int_{k_{-,j}}^{k_{+,j}} \frac{{\rm d}k}{k} {\rm Im}\left[ \frac{-1}{\epsilon(k,W+B_j)} \right]_j \right\} \, \mbox{,}
    \label{eq:SDCSeionis}
\end{equation}
where the limits in the momentum transfer integrals 
are given by:
\begin{equation}
    \hbar k_{\pm,\alpha} = \sqrt{2mT} \pm \sqrt{2m(T-E)} = \sqrt{2mT} \pm \sqrt{2m(T-W-B_{\alpha})} \, \mbox{,}
    \label{eq:limitkFBA}
\end{equation}
with $E=W+B_{\alpha}$, where $\alpha={\rm out}/j$ for the outer/inner-shells.

Further integration in energy of Eq. (\ref{eq:SDCSeionis}) gives the total ionisation cross-section (TICS):
\begin{equation}
    \left.\sigma^{\rm ionis}(T)\right|_{\rm FBA} = 
    \frac{e^2}{\pi \hbar^2 {\cal N}} \frac{m}{T} \left\{ \int_{W_{-,{\rm out}}}^{W_{+,{\rm out}}} {\rm d}W \int_{k_{-,{\rm out}}}^{k_{+,{\rm out}}} \frac{{\rm d}k}{k} {\rm Im}\left[ \frac{-1}{\epsilon(k,W+B_{\rm out})} \right]_{\rm out} 
    + \
    \sum_j \int_{W_{-,j}}^{W_{+,j}} {\rm d}W \int_{k_{-,j}}^{k_{+,j}} \frac{{\rm d}k}{k} {\rm Im}\left[ \frac{-1}{\epsilon(k,W+B_j)} \right]_{j} \right\} \, \mbox{,}
    \label{eq:TICSeFBA}
\end{equation}
and the total electronic excitation cross-section (TECS) is given by:
\begin{equation}
    \left.\sigma^{\rm excit}(T)\right|_{\rm FBA} = \frac{e^2}{\pi \hbar^2 {\cal N}} \frac{m}{T} \int_{E_-}^{E_+} {\rm d}E \int_{k_{-,{\rm out}}}^{k_{+,{\rm out}}} \frac{{\rm d}k}{k} {\rm Im}\left[ \frac{-1}{\epsilon(k,E)} \right]_{\rm out} \, \mbox{,}
    \label{eq:TECSeFBA}
\end{equation}
\end{widetext}
with the integration limits of the momentum transfer given by Eq. (\ref{eq:limitkFBA}). Note that only excitation of the outer shells of the target are considered here.

Let us now examine the integration limits in the energy transfer appearing in Eqs. (\ref{eq:TICSeFBA}) and (\ref{eq:TECSeFBA}). Within the current approximations, electronic excitations can be produced for energy transfers $E$ between the excitation threshold $E_{\rm th}$ and the mean binding energy of the outer shell electrons $B_{\rm out}$, as any larger transfer will lead to ionisation. Note that, when an incident electron has an energy $T<B_{\rm out}$, the maximum energy that it can lose is $T$. Since the primary electron moves in the conduction band and the target electron is promoted to a lower discrete energy level, indistinguishability does not impose any limit to the amount of energy that the primary electron can lose for excitations in the range:
\begin{equation}
    E_- = E_{\rm th} \hspace{10pt} \mbox{,} \hspace{10pt} E_+ = {\rm min}\left[ B_{\rm out}, T \right] \, \mbox{.}
    \label{eq:ExcitLim}
\end{equation}
For ionisation, the limits become:
\begin{eqnarray}
    E_- = B_{\alpha} & \rightarrow & W_{-,\alpha} = 0 \label{eq:ionisLim-} \\
    E_+ = \frac{T+B_{\alpha}}{2} & \rightarrow & W_{+,\alpha} = \frac{T-B_{\alpha}}{2} \, \mbox{.}
    \label{eq:ionisLim+}
\end{eqnarray}
Equation (\ref{eq:ionisLim-}) represents the ionisation threshold, either for outer ($B_{\alpha} = B_{\rm out}$) or inner shells ($B_{\alpha} = B_j$). Equation (\ref{eq:ionisLim+}) puts a limit to the amount of energy that the primary electron can lose, which originates from electron indistinguishability: since now both primary and secondary electrons are indistinguishable particles moving in the conduction band, the primary particle cannot end up with less energy than the secondary electron. In the next section an energy-dependent $B_{\rm out}(T)$ will be introduced, the latter being the actual value entering Eqs. (\ref{eq:limitkFBA}) and (\ref{eq:ExcitLim})-(\ref{eq:ionisLim+}).

\subsection{Mean binding energy for outer-shell electrons}
\label{sec:BmeanT}

The mean binding energy for outer shell electrons can be easily estimated from the ionisation energies for biomolecules reported in the literature, most frequently coming from quantum chemistry calculations \cite{Bernhardt2003, Galassi2012, Bug2017}. In previous works for ion impact \cite{deVera2013, deVera2013b}, $B_{\rm out}$ was calculated as a direct average of the binding energies $B_i$ of the target outer shell electrons, $B_{\rm out}=B_{\rm mean}=\sum_{i=1}^n B_i / n$, with $B_1$ and $B_n$ being, respectively, the first and last outer-shell binding energies (see Table \ref{tab:properties}).  However, it is known that the outermost the electronic shell is, the more it contributes to the ionisation cross-section \cite{Dingfelder1998, Emfietzoglou2007c}, so this fact should be taken into account in some physically motivated manner (especially for low energy electron impact, where shell effects are deemed to be more important \cite{Tan2018}). One of the simpler models for charged-particle impact-ionisation, namely the Rutherford cross-section \cite{ICRU55}, reflects this fact in the sense that the electron binding energy for the $i$-shell appears squared in the denominator. Thus, we will propose a mean binding energy $B_{\rm mean}$ for each material by weighting with $B_i^{-2}$ the relative contribution of each outer-shell ionisation threshold $B_i$:
\begin{equation}
    B_{\rm mean} = \frac{\displaystyle \sum_{i=1}^n \frac{1}{B_i^2}B_i}{\displaystyle \sum_{i=1}^n \frac{1}{B_i^2}} = \frac{\displaystyle \sum_{i=1}^n \frac{1}{B_i}}{\displaystyle \sum_{i=1}^n \frac{1}{B_i^2}} \, \mbox{.}
    \label{eq:Bmean}
\end{equation}
The terms $\frac{1}{B_i^2}$ in the numerator are weights for each individual contribution to $B_{\rm mean}$, while the denominator $\sum_{i=1}^n \frac{1}{B_i^2}$ ensures normalisation (see Table \ref{tab:properties}). When using this criterion for estimating $B_{\rm mean}$, the calculated ionisation SDCS for liquid water presents its maximum value at secondary electron kinetic energies $W\sim 10$ eV, closer to the reference calculations by Emfietzoglou \cite{Emfietzoglou2003a}. Otherwise, the simple average of $B_{\rm mean}$ produces the maximum of the SDCS at energies $W\leq 5$ eV, which is a behaviour closer to water vapour than to the liquid phase \cite{Emfietzoglou2003b}.

As noted by Tan \textit{et al.} \cite{Tan2018}, a constant $B_{\rm mean}$ (independent of the projectile energy $T$) is only valid for electrons with large kinetic energies. When the primary electron has an energy $T \leq B_n$, it is not capable of ionising all the possible electronic levels, and thus we define a new energy-dependent mean binding energy $B_{\rm out}(T)$:
\begin{equation}
B_{\rm out}(T) = \left\{
\begin{tabular}{ll}
$\displaystyle \frac{\sum_{i=1}^{\ell} \frac{1}{B_i}}{\sum_{i=1}^{\ell} \frac{1}{B_i^2}} \, ,$ & $B_1\leq T\leq B_n$ \\
$\displaystyle B_{\rm mean} 
\, ,$ & $T > B_n$ \\
$B_1 \, ,$ & $T < B_1$
\end{tabular}
\right.
\label{eq:BmeanT}
\end{equation}
where 
$B_{\ell}$ is the ionisation threshold immediately below $T$. 
As it will be discussed later, such a correction will notably improve the calculated cross-sections for electrons with energies $T<B_n$. It should be noted that electrons with energies below the first ionisation threshold $B_1$ 
will, in fact, not be able to ionise the medium, and will be only capable of producing electronic excitations with the energy transfer limits given by Eq. (\ref{eq:ExcitLim}).

\begin{figure}[t]
\centering
  \includegraphics[width=\columnwidth]{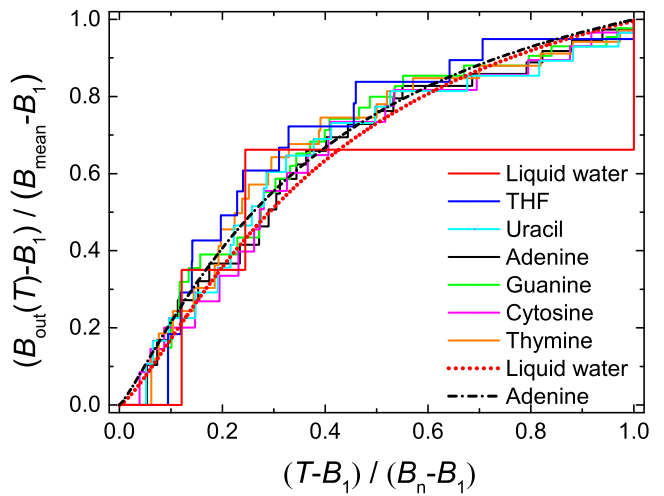}
  \caption{Normalized values of $B_{\rm out}(T)$ as a function of the normalized electron incident energy $T$ for the biological materials studied in this work. Step functions correspond to the values provided by Eq. (\ref{eq:BmeanT}), using quantum chemistry calculations for the binding energies \cite{Bernhardt2003, Galassi2012,Bug2017}, while discontinuous lines correspond to fittings by means of Eq. (\ref{eq:logistic}) for liquid water and adenine.}
  \label{fig:BmeanT}
\end{figure}

Noteworthy, the behaviour of $B_{\rm out}(T)$ is rather similar for the different biomolecules, and can be fitted, for convenience of the numerical implementation, with a logistic function of the form:
\begin{equation}
    B_{\rm out}(T) = B_1 +(B_{\rm mean}-B_1)\left[ a+\frac{b-a}{1+\left( \displaystyle \frac{ \frac{T-B_1}{B_n-B_1}}{c} \right)^d} \right] \, \mbox{,}
    \label{eq:logistic}
\end{equation}
with $a$, $b$, $c$ and $d$ being fitting parameters (given in Table \ref{tab:logistic}). Figure \ref{fig:BmeanT} depicts the normalized function ($B_{\rm out}(T)-B_1)/(B_{\rm mean}-B_1$) as a function of the normalized incident energy ($T-B_1)/(B_n-B_1$), for liquid water, condensed THF, uracil adenine, guanine, cytosine, and thymine (whose properties are summarized in Table \ref{tab:properties}). The steps arise because new ionisation channels only become available once the electron energy is larger than the corresponding binding energy. Clearly, all materials follow a rather similar behaviour when using the normalized functions for $B_{\rm out}(T)$ and $T$. Discontinuous lines illustrate the fittings for liquid water and solid adenine by means of Eq. (\ref{eq:logistic}).

\begin{table}[h]
\small
  \caption{\ Fitting parameters for the dependence of $B_{\rm out}$ on electron energy $T$, Eq. (\ref{eq:logistic}), for the different materials studied in this work}
  \label{tab:logistic}
  \begin{tabular*}{0.48\textwidth}{@{\extracolsep{\fill}}lllll}
    \hline
    Material & $a$ & $b \cdot 10^{4}$ & $c$ & $d$ \\
    \hline
    Liquid water & 1.31752 & -6.62797 & 0.42317 & 1.31513 \\
    THF & 1.103 & -1.04493 & 0.22376 & 1.51679 \\
    Uracil & 1.2302 & 1.32923 & 0.32648 & 1.30721 \\
    Adenine & 1.33102 & 0.826314 & 0.39745 & 1.19642 \\
    Guanine & 1.24236 & 0.336896 & 0.32618 & 1.26275 \\
    Cytosine & 1.2662 & 3.69257 & 0.36651 & 1.31606 \\
    Thymine & 1.16425 & -0.170308 & 0.2776 & 1.40323 \\
    \hline
  \end{tabular*}
\end{table}

\subsection{Very low energy electrons: exchange effects and higher-order Born corrections}
\label{sec:lowT}

When the electrons have a relatively low energy, the FBA for the DDCS, Eq. (\ref{eq:FBAions}), fails due to: (i) the exchange effects in the electron-electron interaction, arising from the indistinguishability of the scattered and the ejected electrons, which cannot be neglected since both have now comparable energies, and (ii) the energy loss is now comparable to the incident electron energy, so the scattering cannot be treated within a first order approximation. Therefore, the exchange terms in Eq. (\ref{eq:DDCS1}) cannot be disregarded anymore and the DDCS has to be written as:
\begin{equation}
    \frac{{\rm d}^2 \sigma(T,E,k)}{{\rm d}E {\rm d}k} = \left.\frac{{\rm d}^2 \sigma(T,E,k)}{{\rm d}E {\rm d}k}\right|_{\rm FBA} + \left.\frac{{\rm d}^2 \sigma(T,E,k)}{{\rm d}E {\rm d}k}\right|_{\rm xc} \, \mbox{,}
    \label{eq:DDCSplusxc}
\end{equation}
where \resizebox{0.88\columnwidth}{!}{$\left.\frac{{\rm d}^2 \sigma(T,E,k)}{{\rm d}E {\rm d}k}\right|_{\rm xc} \propto |g(\Vec{k_1},\Vec{k_2})|^2-{\rm Re}[f(\Vec{k_1},\Vec{k_2}) g^*(\Vec{k_1},\Vec{k_2})]$} represents the exchange term. Several methods have been developed to account for this exchange term \cite{Rudge1968, Emfietzoglou2017a}. Ochkur \cite{Ochkur1965} developed a modification of the first-order Born-Oppenheimer perturbation treatment \cite{Oppenheimer1928} by extrapolating the exchange effect for incident energies lower than the original approximation. The underlying argument was that the Born-Oppenheimer expression for the exchange scattering amplitude (Eq.(3-24) in \cite{Rudge1968}) is correct at high energies, asserting that it is better to retain just the leading term in an expansion of $g(\Vec{k_1},\Vec{k_2})$ in powers of $1/k_0$ ($\hbar k_0$ being the momentum of the primary electron before the collision) rather than the full Born-Oppenheimer expression. The Born-Ochkur (BO) approximation gives a convenient expression for the exchange scattering amplitude that retains the FBA component for the direct-scattering amplitude. The BO exchange amplitudes for the excitation \cite{Ochkur1965,Inokuti1967, Rudge1968} and ionisation \cite{Ochkur1965, Prasad1965, Rudge1968} processes with momentum transfer $\Vec{k}=\Vec{k_0}-\Vec{k_1}$ are:
\begin{eqnarray}
    g_{\rm BO}^{\rm excit}(\Vec{k_1},\Vec{k_2}) & = & \frac{k^2}{k_0^2}f_{\rm FBA}(\Vec{k_1},\Vec{k_2}) 
    \, \mbox{,} 
     \\
    g_{\rm BO}^{\rm ionis}(\Vec{k_1},\Vec{k_2}) & = & \frac{k^2}{k_0^2-k_2^2}f_{\rm FBA}(\Vec{k_1},\Vec{k_2}) \, \mbox{.} 
    \label{eq:BOterms}
\end{eqnarray}

From Eqs. (\ref{eq:FBAions}) and (\ref{eq:DDCSplusxc})-(\ref{eq:BOterms}), the following expressions can be written for the exchange contribution to the excitation and ionisation DDCS:
\begin{align}
    &\left.\frac{{\rm d}^2 \sigma^{\rm excit}(T,E,k)}{{\rm d}E {\rm d}k}\right|_{\rm xc}  =  F_{\rm xc}^{\rm excit}(T,k) \left.\frac{{\rm d}^2 \sigma^{\rm excit}(T,E,k)}{{\rm d}E {\rm d}k}\right|_{\rm FBA}  \, \mbox{,} \label{eq:DDCSexcitXC} \\
    &\left.\frac{{\rm d}^2 \sigma^{\rm ionis}(T,W,k)}{{\rm d}W {\rm d}k}\right|_{\rm xc}  =  F_{\rm xc}^{\rm ionis}(T,W,k) \left.\frac{{\rm d}^2 \sigma^{\rm ionis}(T,W,k)}{{\rm d}W {\rm d}k}\right|_{\rm FBA}  \, \mbox{,} \label{eq:DDCSionisXC}
\end{align}
where $F_{\rm xc}^{\rm excit}(T,k) =  -\frac{k^2/2m}{T}+\left(\frac{k^2/2m}{T}\right)^2 $ and $F_{\rm xc}^{\rm ionis}(T,W,k) =  -\frac{k^2/2m}{T-W}+\left(\frac{k^2/2m}{T-W}\right)^2 $ are the BO exchange correction factors to the DDCS$|_{\rm FBA}$ given by Eq. (\ref{eq:FBAions}). It should be noted that the BO exchange term for ionisation, which is based on a quantum mechanical treatment of the problem, reproduces the classical Mott exchange for binary collisions \cite{ICRU55} at high electron energies. A similar expression for ionisation was used by \cite{Fernandez-Varea1993}.

In the case of slow electrons ($T\leq 100$ eV), the energy loss is no longer negligible with respect to the primary electron energy and the FBA becomes inaccurate. There exist several alternatives to correct the FBA at low energies, ranging from heuristic or phenomenological but easily implementable approaches \cite{Emfietzoglou2005}, to more sophisticated theoretical models \cite{Emfietzoglou2017a}. Since in this work we aim at applying the method to arbitrary biological materials, we prefer to implement a simple scheme, which does not need the determination of additional parameters for the target material. This approach consists on accounting for a classical Coulomb-field correction, which considers the potential energy gained by the incident electron in the field of the target molecule. In practice, this means an increase of the kinetic energy $T$ of the incident electron, which is replaced in Eq. (\ref{eq:FBAions}) in the following manner \cite{Emfietzoglou2005}:
\begin{eqnarray}
    T & \rightarrow & T' \simeq T+2E_i \hspace{10pt} \, \hspace{10pt} \mbox{(for excitations) ,} \label{eq:CoulombExcit} \\
    T & \rightarrow & T'' \simeq T+2B_{\rm \alpha} \hspace{10pt} \, \hspace{10pt} \mbox{(for ionisations) ,} \label{eq:Coulombionis}
\end{eqnarray}
where $B_{\alpha}$ is the electron binding energy ($B_{\alpha} = B_{\rm out}(T)$ for the target outer shells and $B_{\alpha} = B_j$ for the inner shells $j$), and $E_i$ is the energy of the $i$-th discrete electronic transition (see Fig. \ref{fig:ELF}(a)). Within our approach, the latter is simplified to the threshold energy for electronic excitations $E_i=E_{\rm th}$, as individual excitation channels are not explicitly considered (see Table \ref{tab:properties}).

The relevant TECS, ionisation SDCS and TICS quantities can now be expressed including low energy (higher-order, HO) corrections to the direct term plus the exchange term. The resulting TECS will be:
\begin{widetext}
\begin{align}
    \sigma^{\rm excit}(T) &= \left. \sigma^{\rm excit}(T)\right|_{\rm HO} + \left. \sigma^{\rm excit}(T)\right|_{\rm xc} \label{eq:TECSfinal} \\
    &
    =  \frac{me^2}{\pi \hbar^2 {\cal N}} \left\{ \frac{1}{T+2E_{\rm th}} \int_{E_-}^{E_+} {\rm d}E \int_{\kappa_-}^{\kappa_+} \frac{{\rm d}k}{k} {\rm Im}\left[ \frac{-1}{\epsilon(k,E)} \right]_{\rm out} 
    + 
    \frac{1}{T} \int_{E_-}^{E_+} {\rm d}E \int_{k_{-,{\rm out}}}^{k_{+,{\rm out}}} \frac{{\rm d}k}{k} F_{\rm xc}^{\rm excit}(T,k) {\rm Im}\left[ \frac{-1}{\epsilon(k,E)} \right]_{\rm out} \right\} \, \mbox{,} \nonumber 
\end{align}
where $E_{\pm}$ are given by Eq. (\ref{eq:ExcitLim}), $\hbar k_{\pm,{\rm out}}$ by Eq. (\ref{eq:limitkFBA}) and the higher-order momentum transfer limits for excitation (due to Eq.(\ref{eq:CoulombExcit})) are given by \cite{Vriens1966}
    $\hbar \kappa_{\pm} = \sqrt{2m(T+2E_{\rm th})} \pm \sqrt{2m(T+2E_{\rm th}-E)}$
.

For ionisation, the SDCS is written as:
\begin{align}
    &\frac{\sigma^{\rm ionis}(T,W)}{{\rm d}W} =  \left. \frac{\sigma^{\rm ionis}(T,W)}{{\rm d}W}\right|_{\rm HO} + \left. \frac{\sigma^{\rm ionis}(T,W)}{{\rm d}W}\right|_{\rm xc} \label{eq:SDCSfinal}  \\
    &= \frac{me^2}{\pi \hbar^2 {\cal N}} \left\{ \frac{1}{T+2B_{\rm out}(T)} \int_{\mbox{\Fontauri\slshape k}_{-,{\rm out}}}^{\mbox{\Fontauri\slshape k}_{+,{\rm out}}} \frac{{\rm d}k}{k} {\rm Im}\left[ \frac{-1}{\epsilon(k,W+B_{\rm out}(T))} \right]_{\rm out} 
    \right. 
    + \sum_j \frac{1}{T+2B_j} \int_{\mbox{\Fontauri\slshape k}_{-,j}}^{\mbox{\Fontauri\slshape k}_{+,j}} \frac{{\rm d}k}{k} {\rm Im}\left[ \frac{-1}{\epsilon(k,W+B_j)} \right]_j \nonumber \\
    & + \frac{1}{T} \int_{k_{-,{\rm out}}}^{k_{+,{\rm out}}} \frac{{\rm d}k}{k} F_{\rm xc}^{\rm ionis}(T,W,k) {\rm Im}\left[ \frac{-1}{\epsilon(k,W+B_{\rm out}(T))} \right]_{\rm out} 
    + \left. \frac{1}{T} \sum_j \int_{k_{-,j}}^{k_{+,j}} \frac{{\rm d}k}{k} F_{\rm xc}^{\rm ionis}(T,W,k) {\rm Im}\left[ \frac{-1}{\epsilon(k,W+B_j)} \right]_j \right\} \, \mbox{,} \nonumber
\end{align}
and the TICS is:
\begin{align}
    &\sigma^{\rm ionis}(T) =  \left. \sigma^{\rm ionis}(T)\right|_{\rm HO} + \left. \sigma^{\rm ionis}(T)\right|_{\rm xc} \label{eq:TICSfinal} \\
    &= \frac{me^2}{\pi \hbar^2 {\cal N}} \left\{ \frac{1}{T+2B_{\rm out}(T)} \int_{W_{-,{\rm out}}}^{W_{+,{\rm out}}} {\rm d}W\, \int_{\mbox{\Fontauri\slshape k}_{-,{\rm out}}}^{\mbox{\Fontauri\slshape k}_{+,{\rm out}}} \frac{{\rm d}k}{k} {\rm Im}\left[ \frac{-1}{\epsilon(k,W+B_{\rm out}(T))} \right]_{\rm out} \right. \nonumber \\ 
    &\qquad {} 
    + \sum_j \frac{1}{T+2B_j} \int_{W_{-,j}}^{W_{+,j}} {\rm d}W\, \int_{\mbox{\Fontauri\slshape k}_{-,j}}^{\mbox{\Fontauri\slshape k}_{+,j}} \frac{{\rm d}k}{k} {\rm Im}\left[ \frac{-1}{\epsilon(k,W+B_j)} \right]_j \nonumber \\
    &\qquad {} + \frac{1}{T} \int_{W_{-,{\rm out}}}^{W_{+,{\rm out}}} {\rm d}W\, \int_{k_{-,{\rm out}}}^{k_{+,{\rm out}}} \frac{{\rm d}k}{k} F_{\rm xc}^{\rm ionis}(T,W,k) {\rm Im}\left[ \frac{-1}{\epsilon(k,W+B_{\rm out}(T))} \right]_{\rm out} \nonumber \\
    &\qquad {} + \left. \frac{1}{T} \sum_j \int_{W_{-,j}}^{W_{+,j}} {\rm d}W\, \int_{k_{-,j}}^{k_{+,j}} \frac{{\rm d}k}{k} F_{\rm xc}^{\rm ionis}(T,W,k) {\rm Im}\left[ \frac{-1}{\epsilon(k,W+B_j)} \right]_j \right\} \, \mbox{,} \nonumber 
\end{align}
\end{widetext}
where $W_{\pm,\alpha}$ are given by Eqs. (\ref{eq:ionisLim-})-(\ref{eq:ionisLim+}), $\hbar k_{\pm,\alpha}$ by Eq. (\ref{eq:limitkFBA}), and the higher-order momentum limits are \cite{Vriens1966}
    $\hbar \mbox{\Fontauri\slshape k}_{\pm,\alpha} = \sqrt{2m(T+2B_{\alpha})} \pm \sqrt{2m\left[(T+2B_{\alpha})-(W+B_{\alpha})\right]}$, 
with $B_{\alpha}=B_{\rm out}(T)$ for the outer shells, Eq. (\ref{eq:BmeanT}), and $B_{\alpha}=B_j$ for the inner shells $j$.


\section{Results and discussion}
\label{sec:results}

In what follows we apply the present methodology to calculate ionisation singly differential (SDCS), total ionisation (TICS) and total excitation (TECS) cross-sections for the impact of electrons, in a wide energy range, on the relevant condensed-phase biological materials, namely liquid water and the DNA/RNA molecular components (THF, cytosine, adenine, guanine, uracil, and thymine), whose properties are summarized in Table \ref{tab:properties}. Our results will be compared to available experimental data for these targets, as well as analysed in the light of various scaling rules with the number of valence electrons. Since experimental information for condensed-phase targets is very scarce, in most cases our condensed-phase calculations will be compared to gas-phase data.

We start by considering liquid water, the most abundant material in living beings, whose ELF is experimentally available for several values of the momentum transfer \cite{Hayashi2000,Watanabe1997} and for which the MELF-GOS method produces excellent results in comparison with experiments for arbitrary momentum transfers \cite{Garcia-Molina2009, Garcia-Molina2011}. Let us begin by analysing how the different improvements of our model affect the calculated energy distribution of secondary electrons (i.e., the SDCS), which is a very relevant quantity affecting the density of inelastic events of the radiation track structure in living tissue, especially at the nanometre scale, and that will also determine the mechanisms by which secondary electrons damage organic molecules solvated in water.\cite{Solovyov2017} In order to compare with the available experimental data for water vapour \cite{Bolorizadeh1986}, it is important to have in mind the differences in the energy spectra of the liquid and gaseous phases, which results in a shift of the position of the maximum in the energy spectrum from $\sim3$ eV (vapour) to $\sim10$ eV (liquid), according to detailed calculations.\cite{Emfietzoglou2003a, Emfietzoglou2003b}

\begin{figure}[t]
\centering
  \includegraphics[width=\columnwidth]{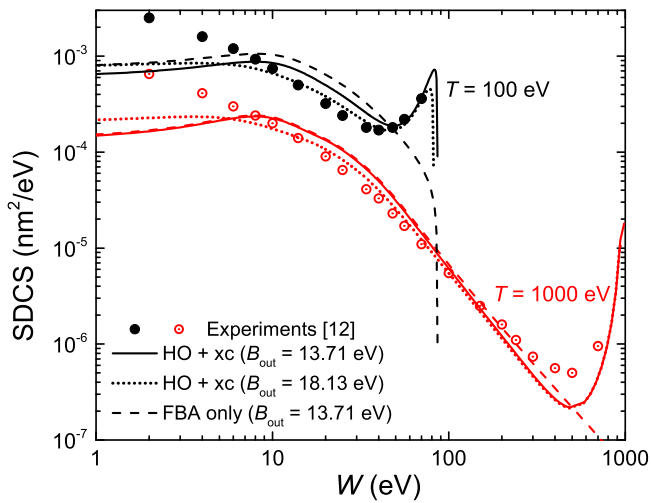}
  \caption{ionisation singly differential cross-sections (SDCS) for 100 and 1000 eV electron impact in water. Symbols represent experimental data for water vapour \cite{Bolorizadeh1986} and lines are our calculations for liquid water. Solid lines represent full calculations (Eq. (\ref{eq:SDCSfinal})) using the value of $B_{\rm mean}$ from Eq. (\ref{eq:BmeanT}), while dotted lines employ a direct average for $B_{\rm mean}$; dashed lines depict the raw FBA calculations 
  (Eq. (\ref{eq:SDCSeionis})).} 
  \label{fig:SDCSwater}
\end{figure}

Figure \ref{fig:SDCSwater} compares the present calculated SDCS (lines) for liquid water with the experimental data (symbols) for water vapour,\cite{Bolorizadeh1986} for $T=100$ and 1000 eV primary electrons. Solid lines depict the results we obtain when using $B_{\rm mean}=13.71$ eV (Eq. (\ref{eq:Bmean})), while dotted lines employ the direct average of outer-shell binding energies introduced in previous works, $B_{\rm mean}=18.13$ eV.\cite{deVera2013, deVera2013b} While both calculations lead to SDCS in a fair agreement with experiments at high energies, it is evident that the weighted $B_{\rm mean}$ gives the maximum of the SDCS at $\sim 10$ eV, in agreement with the predictions for the liquid phase. 
In order to also assess the effect of the exchange (xc) and higher-order (HO) corrections introduced in this work, dashed lines in Fig. \ref{fig:SDCSwater} depict the raw results from the FBA (Eq. (\ref{eq:SDCSeionis})). On the one hand, the exchange correction is needed to yield SDCS that correctly reproduce the primary peak observed at $W\simeq T$, which is due to the indistinguishability between the primary and the secondary electrons. On the other hand, the higher-order correction also diminishes the absolute values of the SDCS for the lower energies $T$, approaching calculations to experimental data (this effect is more clearly seen for $T=100$ eV electrons).

\begin{figure}[t]
\centering
  \includegraphics[width=0.95\columnwidth]{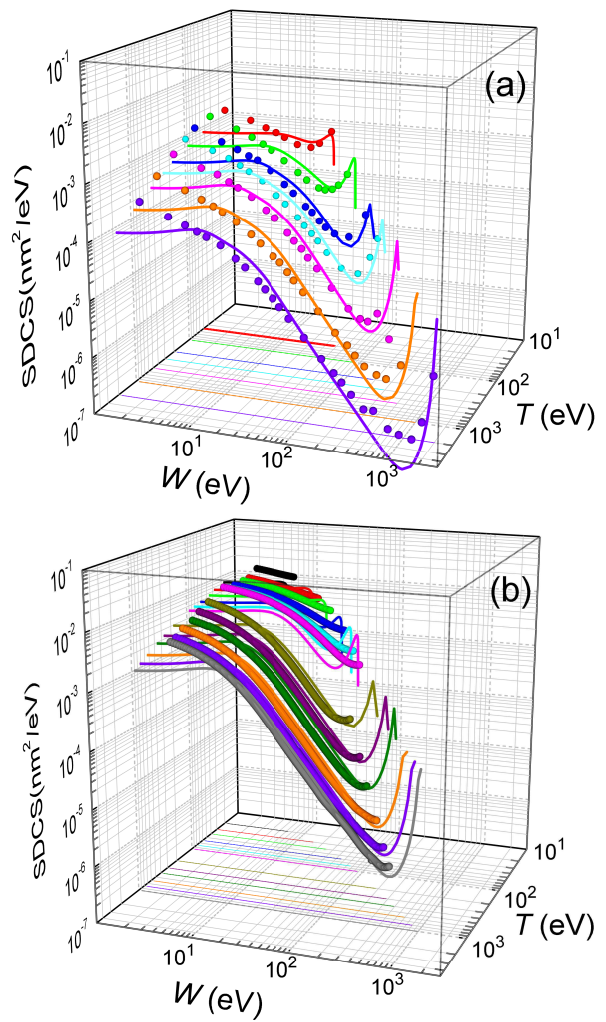}
  \caption{Energy distributions of the electrons generated by the impact of a primary electron (a) with energy $T=$ 50, 100, 200, 300, 500, 1000 and 2000 eV in liquid water, and (b) with energy $T=$ 20, 30, 40, 60, 80, 100, 200, 400, 600, 800 and 1000 eV in THF. Solid lines represent our calculations (Eq. (\ref{eq:SDCSfinal})) and symbols correspond to experimental data for water vapour \cite{Bolorizadeh1986} and THF \cite{Baek2012, Bug2014}. 
  }
  \label{fig:SDCSall}
\end{figure}

Further comparison of the calculated energy distributions of secondary electrons ejected from liquid water for different primary energies $T$ is provided in Fig. \ref{fig:SDCSall}(a), where our full calculations (lines) are compared with all the available experimental data (symbols) for the gas phase \cite{Bolorizadeh1986}. As can be seen, except for the differences expected between the gas and condensed phases (as explained above), there is a very good general agreement of the calculated SDCS with the experimental data, both in shape and absolute values. As the primary electron energy $T$ grows, the height of the SDCS progressively drops, while the high-$W$ tail extends to larger values, as a consequence of the larger primary electron energy and thus to the possibility to eject secondary electrons (or scattered primary electrons) of larger kinetic energies. In any case, the calculated SDCS are lower than the experimental ones for very low electron energies, $W<10$ eV, which might be due to phase effects or to the uncertainty of the measured SDCS at these very low energies.

\begin{figure}[t]
\centering
  \includegraphics[width=\columnwidth]{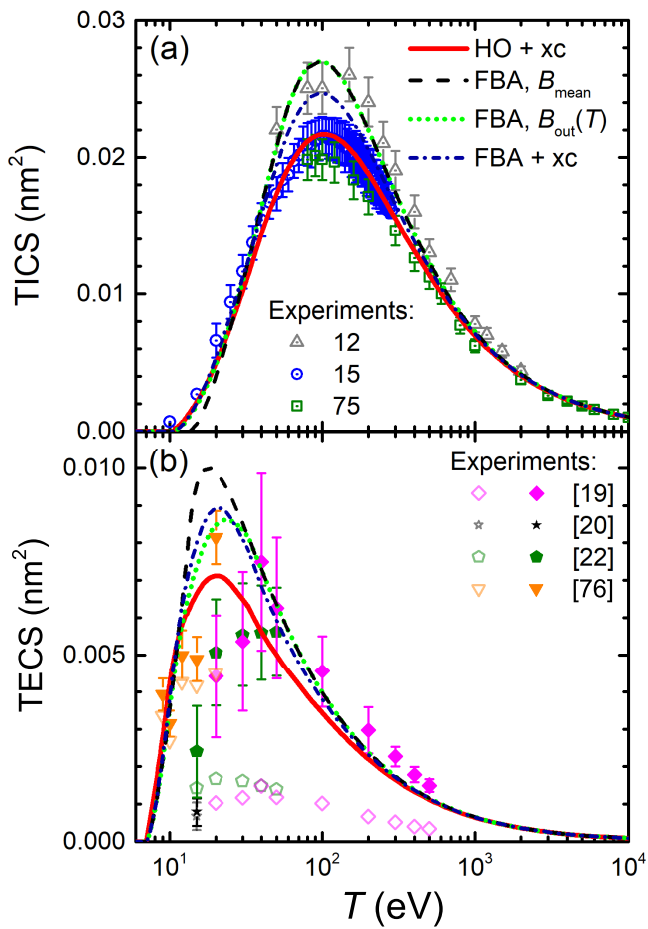}
  \caption{(a) TICS and (b) TECS for electrons in liquid water. Lines represent calculations for the liquid for different levels of approximation, while symbols depict experimental data for water vapour from the different sources indicated in the figure. For excitation, pale open symbols represent raw data obtained by the different authors for a limited number of channels, while full symbols correspond to estimates of the total excitation cross-section, including all possible channels, as explained in the text.}
  \label{fig:TICSTECSwater}
\end{figure}

Plenty of experimental data is available for the total ionisation and excitation cross-sections of water, TICS and TECS, respectively, which allows further assessing the methodology and the different improvements accounted for in this work, as well as to evaluate the relative importance of the electronic ionisation and excitation processes for this relevant material. Figure \ref{fig:TICSTECSwater} shows the calculated and experimental TICS (a) and TECS (b) in liquid water as lines and symbols, respectively. A wealth of experimental data is available for the electron-impact ionisation of water vapour, with a representative selection \cite{Schutten1966a, Bolorizadeh1986, Bull2014} shown in the figure. The dashed line corresponds to the raw FBA calculation, where no improvements are applied (Eq. (\ref{eq:TICSeFBA})), while the dotted line includes the energy-dependent $B_{\rm out}(T)$ defined by Eq. (\ref{eq:BmeanT}). Even though both calculations yield results compatible with the experiments for large energies and even around the maximum, the use of $B_{\rm out}(T)$ clearly improves the results at energies $\leq 20$ eV. While a constant mean binding energy makes the TICS to vanish for energies closer to $B_{\rm mean}=13.71$ eV, the energy-dependent binding energy accounts for low energy electrons being capable to ionise only a fraction of the outer shells, thus producing a progressive decrease of the TICS to zero as the electron energy approaches the first binding energy $B_1=10.79$ eV.

Dash-dotted and solid lines in Fig. \ref{fig:TICSTECSwater}(a) account, respectively, for exchange and for exchange plus higher-order corrections. These improvements have the effect of lowering the TICS, especially around its maximum at $T\sim 100$ eV. When both corrections are accounted for, the calculated TICS for liquid water drops towards the lower limit given by the scattered experimental data for the vapour around the maximum, remarkably coinciding with the most recent experimental data \cite{Bull2014}. This result is to some extent expected, as TICS for liquid water should be lower than for the gas phase, as a consequence of the dielectric screening effect provided by the electron density in the condensed phase.

The results for excitation of water, appearing in Fig. \ref{fig:TICSTECSwater}(b), require further explanations regarding
the available experimental data (again for the gas phase). Measurements of electronic excitation cross-sections are usually restricted to some particular channels. For example, Thorn \textit{et al.} \cite{Thorn2007} only report data for the Ã$^1$B$_1$ excitation, while Matsui \textit{et al.} \cite{Matsui2016} also include the ã$^3$B$_1$ transition, and Brunger \textit{et al.} \cite{Brunger2008} and Ralphs \textit{et al.} \cite{Ralphs2013} present results for the six lowest lying states $^3$B$_1$, $^1$B$_1$, $^3$A$_2$, $^1$A$_2$, $^3$A$_1$, and $^1$A$_1$. These data are presented in Fig. \ref{fig:TICSTECSwater}(b) by pale open symbols. However, data for 19 more transitions are provided in Ref. \cite{Thorn2008}, 
which is the most complete information currently available to the best of our knowledge. Since that work contains all the channels reported by the aforementioned authors, it is possible to use the (more complete) latter data to scale the former partial values as if all these channels were comprised, as our calculated TECS include all possible channels. Scaling factors have been estimated from Thorn’s experiments \cite{Thorn2008} as the ratio of the data for all 25 measured channels to the data for the specific channels at each particular electron energy. Some of the points from \cite{Ralphs2013} and \cite{Thorn2007} lie outside the energy range covered by \cite{Thorn2008} (15-50 eV). For the data from \cite{Ralphs2013}, the ratio at 15 eV has been taken below this energy. For \cite{Thorn2007}, whose data lie at energies larger than 50 eV, an average of the ratios for the available energies between 15 and 50 eV has been used. Error bars correspond to the absolute uncertainties reported by the different authors, when given, or by the relative uncertainties applied to the scaled data. This scaling was done previously \cite{deVera2019} accounting only for 
6 channels ($^3$B$_1$, $^1$B$_1$, $^3$A$_2$, $^1$A$_2$, $^3$A$_1$, and $^1$A$_1$) and not the 25 reported by \cite{Thorn2008}, and hence this is the reason for the difference between the data reported there \cite{deVera2019} with respect to the present work.

The meaning of the lines in Fig. \ref{fig:TICSTECSwater}(b) is the same as in Fig. \ref{fig:TICSTECSwater}(a). The raw FBA calculations using a constant $B_{\rm out}$ (dashed line) already give a reasonable estimate for the TECS in comparison with the scaled experimental data. However, introduction of $B_{\rm out}(T)$ (dotted line) improves the calculated TECS, even more when both exchange and higher-order corrections are also included (full line). Remarkably, the full calculation agrees rather well with the compilation of the (scaled) experimental data, having into account the dispersion of the different datasets and the considerable error bars. Thus, we deem that our calculation gives a reasonable estimate for the total excitation cross-section in water by electron impact. In any case, it is clear that electronic excitations only dominate the inelastic cross-section for energies $<25$ eV, with ionisation being the dominant process for larger electron energies.

The developed methodology allows us to obtain ionisation and excitation cross-sections for any condensed-phase biological material for which the atomic composition and density are known, irrespectively of whether its optical ELF has been experimentally obtained or not. For most of the materials studied in this work the experimental optical ELF is available (see Table \ref{tab:properties}), although there are no measurements for the two relevant DNA molecular components THF and cytosine, whose ELF has been obtained by the parametric predictive model \cite{Tan2004}. 

\begin{figure}[t]
\centering
  \includegraphics[width=\columnwidth]{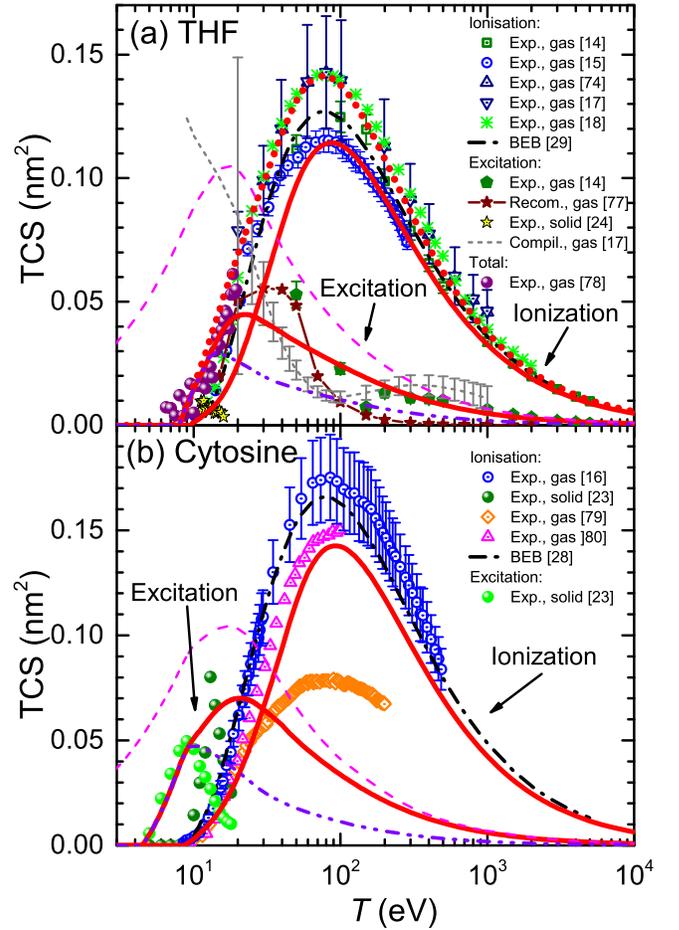}
  \caption{Total electronic cross-section (TCS) for electrons in (a) THF and (b) cytosine, as a function of the electron incident energy $T$. Solid lines correspond to our full calculations for TICS and TECS, while the dotted line in (a) corresponds to the sum of both. Other lines and symbols represent experimental data: for THF, gas-phase TICS \cite{Fuss2009,Bug2014, Bull2014,Bug2017,Wolff2019} and TECS \cite{Fuss2009,Fuss2014, Bug2017}, solid-phase TECS (for $T<11.5$ eV) \cite{Lemelin2016}, and gas-phase total electronic cross-section \cite{Chiari2013}; for cytosine, gas-phase TICS \cite{Shafranyosh2006, Michaud2012, VanderBurgt2014b,Rahman2016} and solid phase TECS \cite{Michaud2012}. BEB calculations for THF \cite{Mozejko2005} and cytosine \cite{Mozejko2003} are depicted by dash-dotted lines. Results from the present model are presented under different assumptions: dashed lines show calculations without using an excitation threshold (i.e., $E_{\rm th}=0$ eV), while dash-dot-dotted lines depict results restricting the excitation process up to 11.5 eV for THF and up to 9 eV for cytosine (see main text for details).}
  \label{fig:TICSTECS_THF_cytosine}
\end{figure}

For the case of THF, our calculated ionisation SDCS by electron impact are compared to available measurements \cite{Baek2012, Bug2014} in Fig. \ref{fig:SDCSall}(b). As in the case of liquid water, the agreement between calculations and experimental data is rather good for the whole energy range of the emitted electrons, except for the differences expected between the condensed and gas phases, which also include, in this case, an underestimation of the SDCS for energies $W\leq 100$ eV. As for the total cross-sections, full calculations (solid lines) are presented in Fig. \ref{fig:TICSTECS_THF_cytosine}(a) along with the available experimental data (symbols) and semiempirical binary-encounter Bethe (BEB) calculations \cite{Mozejko2005} (dash-dotted line). For the TICS, experiments are available only for the gas phase \cite{Fuss2009, Bug2014,Bull2014,Bug2017, Wolff2019}. Our calculations agree well with most of the experiments at large energies and, around the maximum, they are very close to the data by \cite{Bull2014}, the lowest among all experimental results. The differences with respect to the BEB calculations and other experiments can in principle be attributed to phase effects. As for the TECS, full calculated results (solid line) are compared to the experimental data for the condensed phase \cite{Lemelin2016} and for the gas phase \cite{Fuss2009, Bug2017}, as well as the recommended data by \cite{Fuss2014}. Data from \cite{Lemelin2016} are too low in comparison with our calculations, being the agreement rather good with \cite{Fuss2009,Fuss2014,Bug2017}. Although our calculations for energies below 30 eV are quite smaller than the data compiled in \cite{Bug2017}, it should be noted that the latter also includes, apart from electronic excitations, vibrational and rotational excitations, which we did not considered in our treatment. Notably, the calculated total (ionisation plus excitation) electronic cross-section (dotted line) agrees very well with the experimental data (for the gas phase) from \cite{Chiari2013}.

Results for the TICS and TECS for cytosine are presented in Fig. \ref{fig:TICSTECS_THF_cytosine}(b). The former are in fair agreement with the experimental data from \cite{VanderBurgt2014a}, although they are lower than the data from \cite{Rahman2016} and much larger than the results from \cite{Shafranyosh2006}. Having into account the dispersion in the experimental data, the relatively good agreement between the data from \cite{VanderBurgt2014a} and \cite{Rahman2016}, as well as with the BEB calculations from \cite{Mozejko2003}, makes these values plausible for the gas phase. Our calculations for solid cytosine are rather close to these data, although slightly below, most likely due to phase effects. As for the calculated TECS, they correspond very well with the data for the solid form of cytosine \cite{Michaud2012} up to 10-12 eV. A possible reason for the discrepancies for the larger energies is discussed below, in the light of the comparisons for other DNA bases.

The parametric model for predicting the optical ELF of these biomaterials \cite{Tan2004} does not give any information on the energy threshold for electronic excitations, $E_{\rm th}$. For THF and cytosine, these values have been obtained from other sources (see Table \ref{tab:properties}). In order to check the sensitivity of the calculated TECS on the value of $E_{\rm th}$, in Fig. \ref{fig:TICSTECS_THF_cytosine} we show by dashed lines the cross-sections obtained when setting $E_{\rm th}=0$ eV. As can be seen, these values are too large in comparison with the available experimental data, which remarks the importance of counting on with a reliable source for the energy threshold for electronic excitations in order to calculate the TECS (note that the excitation threshold does not affect the TICS).

\begin{figure*}[t]
\centering
  \includegraphics[width=0.8\textwidth]{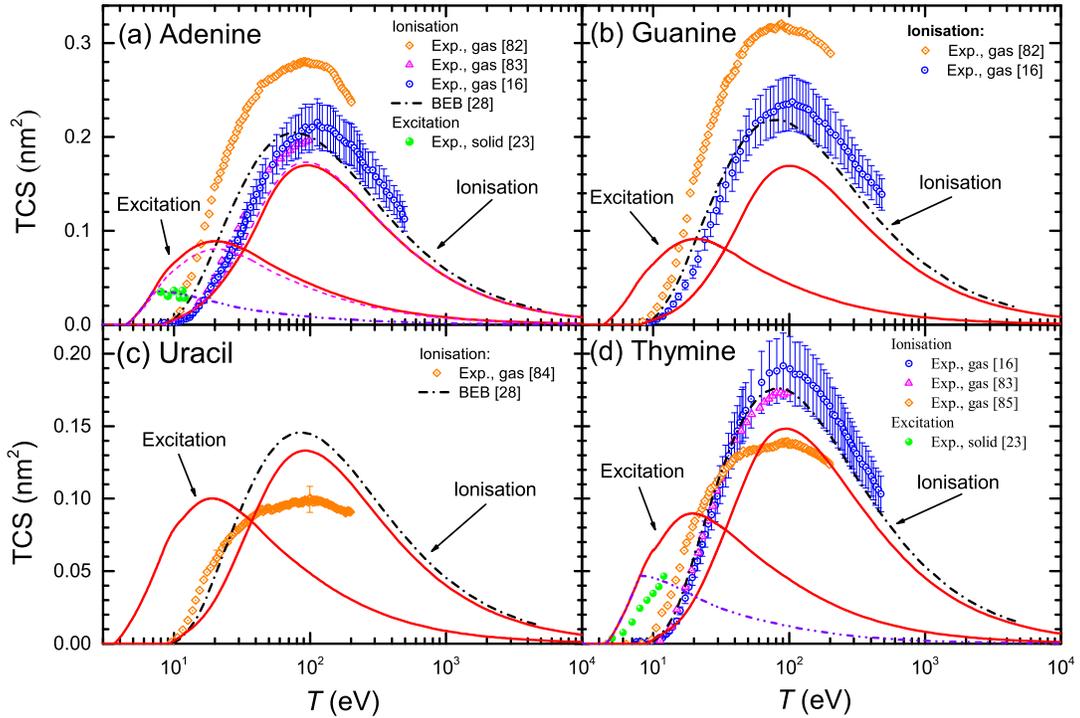}
  \caption{Total electronic cross-sections (TCS) for electrons in the DNA/RNA bases (a) adenine, (b) guanine, (c) uracil and (d) thymine, as a function of the incident energy $T$ for excitation and ionisation processes. Solid lines are present full calculations. Symbols correspond to experimental data: for electron ionisation in adenine (gas) \cite{Minaev2014, VanderBurgt2015, Rahman2016}; in guanine (gas) \cite{Minaev2014, Rahman2016}; in uracil (gas) \cite{Shafranyosh2012}; in thymine (gas) \cite{Shafranyosh2008, VanderBurgt2015, Rahman2016}. For excitation processes: in solid adenine and thymine \cite{Michaud2012}. BEB calculations (dash-dotted lines) are from \cite{Mozejko2003}. Further calculations using the present model are presented under different assumptions: long-dashed lines (for adenine) show calculations using the optical ELF from the parametric model \cite{Tan2004} (with threshold $E_{\rm th}$), while short-dashed lines depict results integrating Eq. (\ref{eq:TECSfinal}) up to 7.5 eV for adenine and up to 8 eV for thymine.}
  \label{fig:TICSTECS_bases}
\end{figure*}

The above results for THF and cytosine show the reliability of the methodology, even when the optical ELF of the material is not experimentally available. However, in principle, results are expected to be more accurate when this information is at hand, as it is the case for the other DNA/RNA bases (adenine, guanine, thymine, and uracil), whose calculated TICS and TECS are shown by solid lines in Fig. \ref{fig:TICSTECS_bases}. For adenine (panel (a)), experimental TICS for the gas phase \cite{Minaev2014, VanderBurgt2015, Rahman2016} and TECS for the solid phase \cite{Michaud2012} are available (symbols). The calculated TICS is just slightly below the experiments from \cite{VanderBurgt2015, Rahman2016} and the BEB calculations from \cite{Mozejko2003} (which agree well among themselves), and rather lower than the results from \cite{Minaev2014}. Such slight differences are, again, attributed to phase effects. Regarding the TECS, calculations give results in the same order of magnitude as the experiments in the condensed phase \cite{Michaud2012}, although with differences in shape and intensity that will be discussed in more details latter. Panels (b), (c) and (d) show the results for guanine, uracil and thymine, respectively. As for adenine, experimental TICS (in the gas phase) from \cite{VanderBurgt2015, Rahman2016} and the BEB calculations from \cite{Mozejko2003} agree among themselves, with our results being slightly below, which we attribute to phase effects.  In general, the experimental results from the Uzhgorod's group \cite{Shafranyosh2006,Shafranyosh2008,Shafranyosh2012,Minaev2014} are rather different from data by other experimentalists although, for the particular cases of uracil\cite{Shafranyosh2012} and thymine\cite{Shafranyosh2008}, they give the maximum of the TICS at similar energies and close intensities as our calculations (these results being the only available for uracil). TECS are only experimentally available for solid thymine \cite{Michaud2012}, which presents its onset at a similar energy as our calculations. Even though the slope of the experiments is slightly lower than in our calculations, the agreement is still fair.

In order to further benchmark the performance of the predictive model for the optical ELF \cite{Tan2004}, dashed lines in Fig. \ref{fig:TICSTECS_bases}(a) depict the calculated TICS and TECS for electrons in adenine when using this parametric ELF, instead of the one experimentally measured. As can be seen, this option yields results which are very close to the calculations employing the experimental optical ELF, what gives confidence in the use of the model for predicting electron-impact cross-sections in arbitrary biomaterials for which information is not available, including complex biotargets such as DNA, proteins \cite{deVera2013}, or even cell compartments \cite{deVera2014}.

One more clarification can be done regarding the comparison between the calculated and experimental TECS for THF (Fig. \ref{fig:TICSTECS_THF_cytosine}(a)), cytosine (Fig. \ref{fig:TICSTECS_THF_cytosine}(b)), adenine (Fig. \ref{fig:TICSTECS_bases}(a)) and thymine (Fig. \ref{fig:TICSTECS_bases}(d)). For these materials, measurements are available for the condensed phases from electron energy loss spectroscopy \cite{Michaud2012, Lemelin2016}. These data are very relevant, since they are among the few available for the electron cross-sections determination in solid biomaterials. However, it has to be taken into account that the energy transfer range analysed in these experiments was limited to relatively low values. The experimental electron energy-loss spectrum (EELS) for THF was analysed up to 11.5 eV, for cytosine up to $\sim 9$ eV, for adenine up to $\sim 7$ eV, and for thymine up to $\sim 8.5$ eV. Hence, there is a possibility that not all the electronic excitation channels are probed in the measured TECS, what might justify some discrepancies with our calculations. To check this possibility, we performed further calculations for these materials, artificially establishing an upper limit in the energy transfer given by $E_+={\rm min}[B_{\rm out},T,E_{\rm max}]$ (instead of Eq. (\ref{eq:ExcitLim})), where $E_{\rm max}$ are the maximum energies analysed for electronic excitations in the experimental electron energy loss spectra. The results are shown in Figs. \ref{fig:TICSTECS_THF_cytosine}(a), \ref{fig:TICSTECS_THF_cytosine}(b), \ref{fig:TICSTECS_bases}(a) and \ref{fig:TICSTECS_bases}(d) by dash-dot-dotted lines. It can be seen that, especially for the cases of cytosine and adenine, these calculations resemble much more the experimental data: for cytosine, they give the maximum of the TECS at the same energy and with the same height as the experiments, although with a more progressive decrease; for adenine, the calculations with the limited energy loss range agree almost perfectly with the experimental data; for thymine, the agreement is less remarkable, with the position of the maximum shifted to slightly lower energies in the calculations, although with the maximum value similar to experiments; however, no agreement is found for the case of THF. These findings indicate that our methodology for calculating the TECS for electron impact in biomaterials presents a very reasonable reliability, giving fair results for most of the targets for which there are measurements, either for the gas or the condensed phases.

Now that we count on with a collection of calculated excitation and ionisation cross-sections for electron impact in a wide variety of relevant biomaterials, we can further analyse the obtained results by testing the scaling 
rules for the ion-impact ionisation cross-sections of biomolecules \cite{Wilson1975, Lynch1976}, proposing that the TICS were proportional to the number of valence electrons per molecule. Such rules have been shown useful for analysing and compiling electron and ion cross-sections for biomolecules \cite{Wang2016, Bug2017}. However, recently it has been pointed out that the number of valence electrons may not be the proper quantity for scaling the TICS \cite{Mendez2020}, which instead should be better scaled as a function of the effective number of molecular electrons (i.e., those participating in the ionising collision). In the following, we will analyse both (the older and the new) scaling rules in the light of our calculated cross-sections for electron impact, not only for ionisation, but also for electronic excitation.

\begin{figure*}[t]
\centering
  \includegraphics[width=0.8\textwidth]{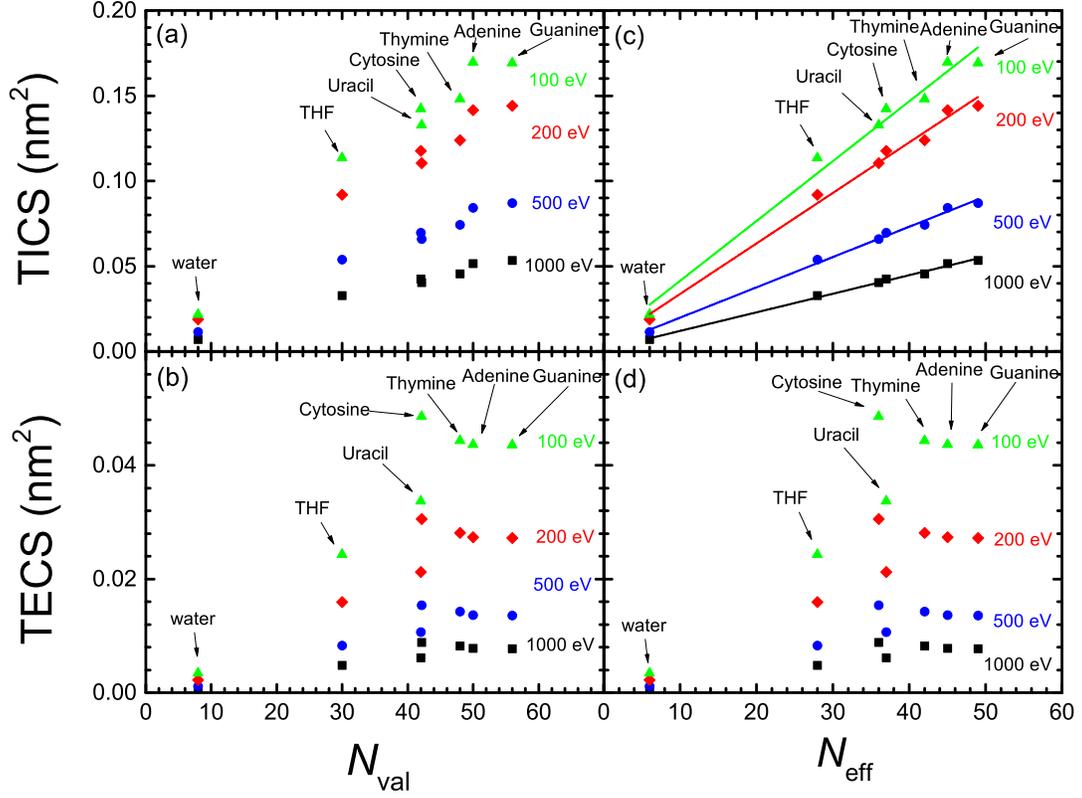}
  \caption{Dependence of the calculated (a) TICS and (b) TECS for biomaterials, for different electron energies ($T=$ 100 eV, 200 eV, 500 eV and 1000 eV), as a function of the number of valence electrons for each biomolecule \cite{Wilson1975, Lynch1976}. Panels (c) and (d) represent the same as (a) and (b), but as a function of the effective number of valence electrons \cite{Mendez2020}. The straight lines in panel (c) represent the linear fittings by means of Eq. (\ref{eq:scaling}) with the parameters given in Table \ref{tab:scaling}.}
  \label{fig:scaling}
\end{figure*}

Figures \ref{fig:scaling}(a) and (b) depict, respectively, the calculated TICS and TECS for the different biomaterials as a function of the number of valence electrons $N_{\rm val}$ (see Table \ref{tab:properties}) for several electron energies. As can be seen, neither TICS nor TECS scale well with the number of valence electrons $N_{\rm val}$. The TECS present a rather more erratic behaviour, and somewhat saturate for $N_{\rm val}\geq 45$, and with cytosine presenting an unusually large TECS. Figures \ref{fig:scaling}(c) and (d) also show, respectively, the TICS and TECS for the different molecules and electron energies, but now as a function of the effective number of electrons $N_{\rm eff}$ proposed by \cite{Mendez2020} (see Table \ref{tab:properties}). Clearly, the linear scaling of the TICS is now better than as a function of $N_{\rm val}$. However, the behaviour of the TECS is not improved at all.

From the above discussion, we can conclude that the proposed scaling laws for ion impact ionisation work well for the electron-impact TICS too. This is true, in principle, for the entire electron energy range, although the linear scaling works better for the larger electron energies. Also, the scaling laws seem more reliable when expressed as a function of the effective number of electrons\cite{Mendez2020} (instead of the number of valence electrons), in the form:
\begin{equation}
    {\rm TICS(nm}^2) = \beta\, N_{\rm eff} + \gamma \, \mbox{,}    \label{eq:scaling}
\end{equation}
where the values of the parameters $\beta$ and $\gamma$ are given in Table \ref{tab:scaling}. Nonetheless, the scaling in the TECS is not satisfactory, neither as a function of the number of valence nor of effective electrons. However, this deficiency does not seem to arise from limitations of the model, as the calculated TECS have been shown reasonably accurate in comparison with experimental data for water and THF vapour, as well as for solid cytosine, adenine and thymine. Thus, we attribute the less predictable behaviour of the TECS to the larger number of variables on which it depends, namely the features of the ELF (related to the number of valence electrons), but also to the values of both the electronic excitation threshold $E_{\rm th}$ and the mean binding energy of outer-shell electrons $B_{\rm out}(T)$, whose values cannot be, in principle, anticipated. 

\begin{table}[b]
\small
  \caption{\ Linear fitting parameters for the dependence of the TICS on the effective number of valence electrons, $N_{\rm eff}$, Eq. (\ref{eq:scaling}), for different initial electron energies $T$}
  \label{tab:scaling}
  \begin{tabular*}{0.48\textwidth}{@{\extracolsep{\fill}}lll}
    \hline
    $T$ (eV) & $\beta$ ($10^3$ nm$^2$/electron) & $\gamma$ ($10^3$ nm$^2$) \\
    \hline
    100 & 3.5 & 6.3 \\
    200 & 3.0 & 4.0 \\
    500 & 1.8 & 2.0 \\
    1000 & 1.1 & 1.1 \\
    \hline
  \end{tabular*}
\end{table}

Finally, to complete our analysis of electronic excitation of biomaterials by electron impact, we provide another important quantity for assessing radiation biodamage, namely the electronic stopping cross-section (SCS) due to both electronic excitations and ionisations SCS = SCS$^{\rm excit} + $SCS$^{\rm ionis}$. The quantities SCS$^{\rm excit}$ and SCS$^{\rm ionis}$ are obtained by integrals similar to Eqs. (\ref{eq:TECSfinal}) and (\ref{eq:TICSfinal}) but replacing d$E$ and d$W$ by d$E\,E$ and d$W\,(W+B_{\alpha})$, respectively.
The SCS gives a measure of the probability of energy deposition in a target molecule. This quantity was suggested \cite{Rezaee2014} as a convenient physical parameter for nanodosimetry studies, as it can be correlated with the damaging cross-sections for double-strand break formation in DNA. 

\begin{figure}[t]
\centering
  \includegraphics[width=\columnwidth]{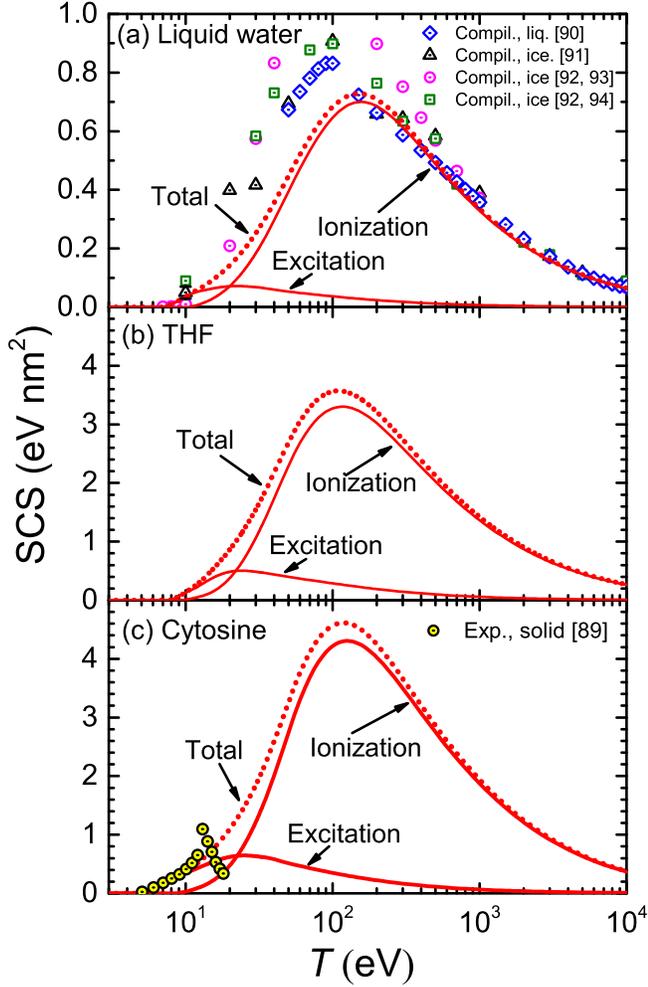}
  \caption{Calculated stopping cross-section SCS (lines) for electrons in (a) liquid water, (b) THF and (c) cytosine due to ionisation and electronic excitation. Total SCS (ionisation+ionisation) are represented by dotted lines.
  Symbols in the upper panel correspond to reference data for liquid water \cite{Watt1996}, ice \cite{ICRU16,LaVerne1985,Luo1991,Joy1995} 
  and cytosine  \cite{Rezaee2014}.}
  \label{fig:SCS1}
\end{figure}

\begin{figure*}[t]
\centering
  \includegraphics[width=0.8\textwidth]{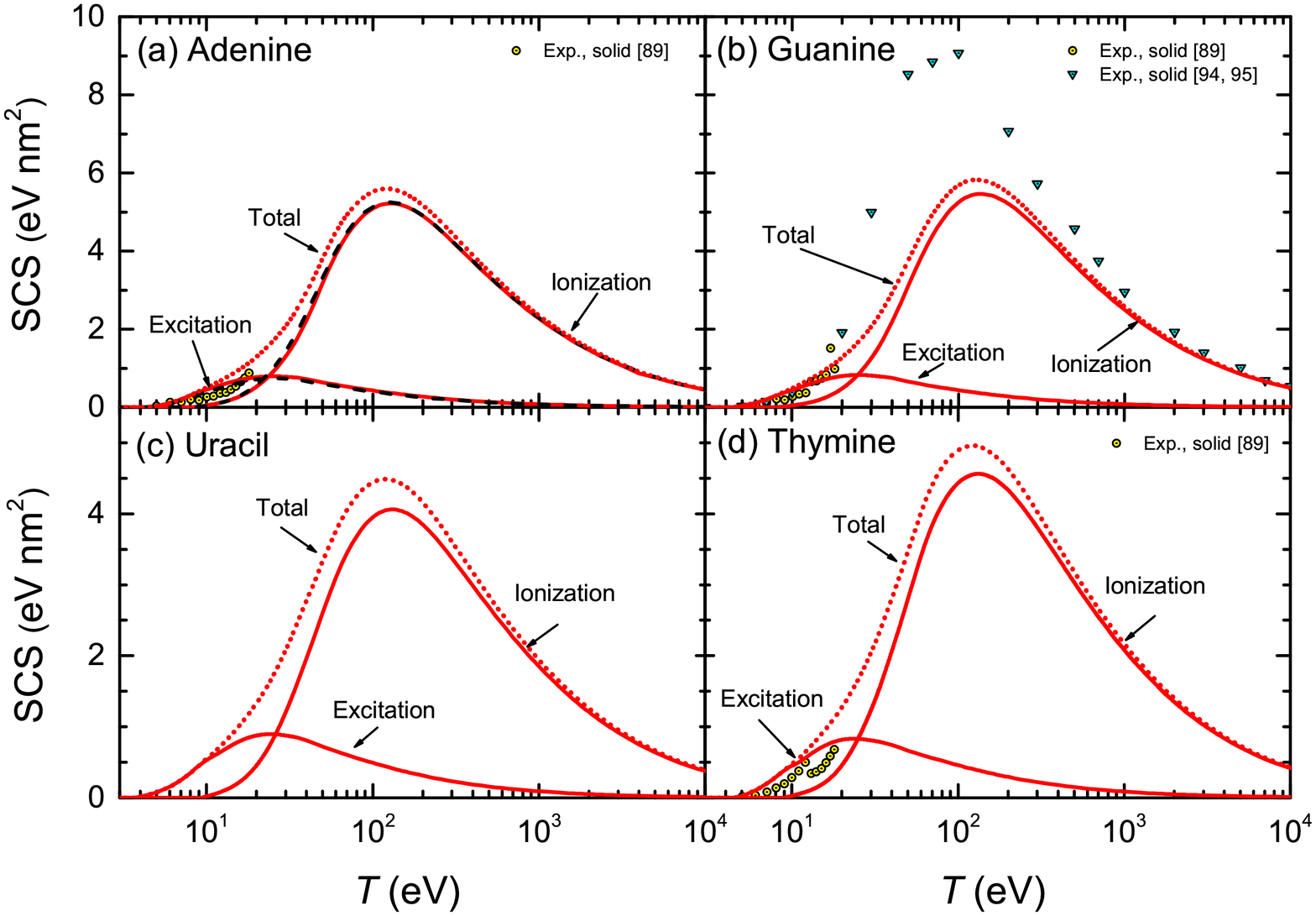}
  \caption{Calculated stopping cross-sections SCS (lines) for electrons in the solid DNA/RNA bases: (a) adenine, (b) guanine, (c) uracil and (d) thymine due to the processes of ionisation and electronic excitation. Total SCS (ionisation+excitation) are represented by dotted lines.
  For adenine, results obtained by using the parametric model for the optical ELF\cite{Tan2004} are presented by dashed lines.
  Symbols represent reference data \cite{Rezaee2014, Luo1994PhD,Joy1995}}
  \label{fig:SCS2}
\end{figure*}

The calculated SCS for all the relevant biomaterials are presented in Fig. \ref{fig:SCS1} (liquid water, THF and cytosine) and Fig. \ref{fig:SCS2} (adenine, guanine, uracil, and thymine). For water (Fig. \ref{fig:SCS1}(a)), calculations are compared with compilations from various sources for liquid water \cite{ICRU16, Watt1996} and ice \cite{LaVerne1985, Luo1991} (Refs. \cite{LaVerne1985,Luo1991} come from Joy's compilation \cite{Joy1995}). Our SCS is split into excitation and ionisation contributions, although comparison with other results can be only done for the total electronic SCS, which agrees well with \cite{Watt1996} for large energies down to the maximum ($\sim 150$ eV). Our calculations significantly underestimate the data for the lower energies \cite{ICRU16, Watt1996, LaVerne1985, Luo1991}. There are no data available for the comparison with THF (Fig. \ref{fig:SCS1}(b)).

The calculated SCS for all the RNA/DNA bases (uracil, adenine, guanine,cytosine, and thymine) appear in Fig. \ref{fig:SCS1}(c) and Fig. \ref{fig:SCS2}. 
There are data for the four DNA bases at low energies ($T\leq 18$ eV) \cite{Rezaee2014}, although it is not clear whether they are experiments or calculations from the reading of the cited article. Our calculated total SCS agree nicely with this data. Only guanine has available SCS data in a higher energy range (20-10$^4$ eV) \cite{Luo1994PhD,Joy1995}. It is worth noting that Luo's SCS for liquid water\cite{Luo1991} and solid guanine\cite{Luo1994PhD} are not a direct measurement, but they come from the experimental determination of the EELS for these materials. 


\section{Summary and conclusions}
\label{sec:concl}

We have extended the dielectric response model, previously developed for the ion-impact ionisation of arbitrary condensed-phase organic media \cite{deVera2013,deVera2013b}, to the case of excitations and ionisations induced by electrons in a wide energy range, due to their relevance in the direct damage of biological targets by different electronic excitations and ionising collisions. Especial attention has been paid to the proper description of very low energy electrons, as most of the secondary electrons produced by radiation have energies below 100 eV. These are nowadays recognized as the most important agents in radiation biodamage, since they are responsible for the development of the radiation track-structure (especially in the nanometer scale, typical of important biomolecules such as DNA).

Such extension to incident electrons required the consideration of a series of improvements to the original model, namely: (i) accounting for the indistinguishability between the primary and the secondary electrons (which affects the range of possible energy transfers in the inelastic collisions), (ii) modification of the concept of an energy-independent mean binding energy for the target’s outer-shell electrons when the primary particle has very low energy (in such a case, the projectile cannot ionise all the target’s outer shells), (iii) inclusion of exchange interaction between the primary and scattered electrons (for both electronic excitations and ionisations), and (iv) introduction of higher-order corrections to the first Born approximation (FBA, on which the dielectric formalism is based) when the energy transfer is comparable to the primary electron energy. All these effects have been incorporated into our model, defining an energy-dependent mean binding energy for the outer-shell electrons on the basis of quantum chemistry data (which show a rather universal behaviour for all analysed materials), accounting for exchange factors for excitation and ionisation processes from the quantum-mechanical Born-Ochkur approximation, as well as implementing a simple higher-order Coulomb-field correction to the FBA, all the above allowing straightforward calculations of single differential and total cross-sections for arbitrary condensed-phase biomaterials.

As a result of such improvements, we have produced a comprehensive set of excitation and ionisation cross-sections for a wide set of biomolecules, including the main constituent of biological tissue, liquid water, as well as the relevant molecular components of DNA/RNA, namely the analogue molecule of the sugar component of the phosphate-deoxyribose backbone, THF, and all the bases adenine, guanine, cytosine, thymine and uracil. Such data comprises the well-studied total ionisation cross-sections (TICS), but also the energy distributions of secondary electrons (singly differential cross-sections, SDCS) as well as the total excitation cross-sections (TECS), for which available data are much scarcer. Moreover, we have calculated the stopping cross-sections (SCS), also related to biomolecular damage. This information is provided over a very wide electron energy range, which covers the high energies characteristic of electron beams, photoelectrons or delta-electrons produced by ion impact, as well as for electrons with kinetic energies down to electronic excitation threshold, so important in molecular damage, constituting by itself a comprehensive and important compilation of cross-sections.

The calculated SDCS, TICS and TECS have been compared with available experimental data that are limited to molecules in the gas phase, except for TECS in DNA bases. Particularly, plenty of experimental data is available for the SDCS and TICS for water and THF, the most studied molecules. The agreement between the model calculations and these experiments is remarkable, having into account the expected differences between the condensed and gas phases that, in this case, are mainly reflected in the position of the maximum of the energy spectra of secondary electrons, located at $\sim 10$ eV for the condensed phase. Regarding their TECS, a good amount of data is available for water, this information being rather more limited for THF. In any case, the calculated TECS also agree remarkably well for liquid water, and give a rather good estimate for THF in comparison with the available reference and recommended data.

As for the DNA/RNA bases, certain underestimation of the calculated TICS around their maxima is observed for some of them, which can be attributed, most likely, to the expected phase effects: in general, the cross-sections for the condensed phase are expected to be lower than for the gas phase, due to the dielectric screening in the electron gas in the solid; nonetheless, the generally large dispersion between different sets of experimental data prevents drawing more definitive conclusions. In any case, the shape and positions of the TICS agree well with experiments, and their absolute values follow well the linear scaling law with the number of effective valence electrons recently proposed on the basis of quantum calculations \cite{Mendez2020}, which gives more confidence in the obtained results. Based on the calculations, we have provided parameters for linear equations predicting the TICS for biomaterials on the basis of their effective number of valence electrons.

Regarding the calculated TECS for the DNA bases, there is a general satisfactory agreement with experimental data in the condensed phase for energies below 10 eV. Deviations are identified for larger electron energies. However, it has been shown that, in several of these cases, such discrepancies seem to arise from the limited energy-loss range analysed in the electron energy loss spectroscopy experiments, as calculations limiting the energy-loss values to a similar range yield partial TECS that, at least for cytosine and adenine, and in part also for thymine, reproduce rather well the experimental results. The TECS, however, do not seem to scale well with the number of valence or effective electrons, what seems to arise from their dependence on a greater number of target material parameters apart from the number of electrons, namely the threshold energy for electronic excitations and the mean binding energy for outer-shell electrons. 

Finally, the stopping cross-sections (SCS) due to both excitation and ionisation processes, relevant to evaluate electron-induced molecular damage, have been obtained and compared with the reference data available for liquid water and all the DNA/RNA bases. The agreement is in general rather good with the most recent datasets obtained from experiments and calculations by Rezaee \textit{et al.} for ultra-low energy electrons \cite{Rezaee2014}, although our findings are lower than the compiled values for water and guanine for energies around the maximum.

All in all, having into account the generally good comparison between calculations and experiments and the dispersion inherent to the measured data, the proposed model is deemed to produce very good estimates for the electron excitation and ionisation cross-sections in biomaterials in the whole energy range. Moreover, the current approach delivers comprehensive information on these interactions, which is not limited to the TICS, as it happens with other simple models, but also provides the energy distributions of secondary electrons (SDCS) as well as the TECS, which are very difficult to obtain from other approaches, as far as we know. Very importantly, the model we have presented is relatively simple to implement, allowing the prediction of the electronic interaction cross-sections for organic or biological condensed-phase materials even when their optical electronic excitation spectrum is unknown (either experimentally or theoretically), by means of applying the parametric approach that provides the optical ELF of the material on the basis of its atomic composition and density \cite{Tan2004}, as well as the electronic excitation threshold energy. Thus, the proposed approach provides a very convenient tool for computing the electronic interaction cross-sections for arbitrary organic and biological materials. These not only include water and the RNA/DNA building blocks analysed here, but also complex biological materials such as proteins, lipids, sugars \cite{deVera2013,Garcia-Molina2017a} or even cell compartments \cite{deVera2014}, as well as other materials relevant for technological applications, such as polymeric resists \cite{deVera2011} or organometallic compounds \cite{deVera2020}. Finally, the exchange and higher-order corrections here implemented have been kept purposely simple, for the sake of the generality and usability of the model. Nonetheless, our approach is equally prepared for implementing other more rigorous quantum approaches for these corrections \cite{Emfietzoglou2017a}, as well as for more accurate estimates of the ELF by means of advanced \textit{ab initio} techniques \cite{Nguyen-Truong2017,Azzolini2017}, which might increase its accuracy for very low energies, objectives which are left for future work.

\section*{Acknowledgements}
Financial support was provided by the Spanish Ministerio de Econom\'{i}a y Competitividad and the European Regional Development Fund (Project no. PGC2018-096788-B-I00), by the Fundaci\'{o}n S\'{e}neca (Project no. 19907/GERM/15) and by the Conselleria d'Educaci\'{o}, Investigaci\'{o}, Cultura i Esport de la Generalitat Valenciana (Project no. AICO/2019/070). PdV acknowledges further financial support provided by the Spanish Ministerio de Econom\'{i}a y Competitividad by means of a Juan de la Cierva postdoctoral fellowship (FJCI-2017-32233).





\bibliography{library}
\bibliographystyle{rsc} 

\end{document}